\documentclass[aps,prl,twocolumn,amsmath,amssymb,showpacs,
floatfix]{revtex4}

\usepackage{epsfig}
\usepackage{bm}
\begin{document}


\title{
Entanglement perturbation theory for the elementary excitation in one dimension
}


\author{
Sung Gong Chung and Lihua Wang}
\affiliation{
Department of Physics and Nanotechnology Research and Computation Center,
 Western Michigan University, Kalamazoo, MI 49008-5252, USA}

\date{\today}

\begin{abstract}

The entanglement perturbation theory is developed to calculate the excitation spectrum in one dimension.
Applied to the spin-$\frac{1}{2}$ antiferromagnetic Heisenberg model, it reproduces 
the des Cloiseaux-Pearson Bethe ansatz result.
As for spin-1, the spin-triplet magnon spectrum has been determined
for the first time for the entire Brillouin zone, including the Haldane gap at $k=\pi$.
\end{abstract}

\pacs{71.10.Li, 02.90.+p, 71.10.Fd, 75.10.Jm}

\maketitle

The importance of elementary excitations in condensed matter systems may be best understood in the
superfluid $^4\mathrm{He}$.  The Tisza two-fluid model with the experimentally 
found phonon-roton spectrum explains fundamental
properties of the superfluid $^4\mathrm{He}$ \cite{til}.  Feynman's effort then to explain the roton
spectrum is well known \cite{fey}. From the theorem of Bloch-Floquet, the elementary excitation with momentum $k$
for a translationally invariant Hamiltonian $H$ is written as 
\begin{equation}\label{eq1}
\Psi_k=\sum_{l=1}^N e^{ikl}O_l^{\dagger}|g\rangle
\end{equation}
where $|g\rangle$ is the ground state and the summation over $l$ extends over the entire lattice sites.
The $O_l$ is a local cluster operator to be determined for a given Hamiltonian.

In spite of a simplicity and validity of the expression (\ref{eq1}), not much progress
has been made along this line since the days of Feynman.  The Heisenberg 
antiferromagnet (HA) described by the Hamiltonian
\begin{equation}\label{eq2}
H=J\sum_i \textbf{S}_i\cdot\textbf{S}_{i+1}
\end{equation}
is probably the best studied system concerning the excitation spectrum.  In particular,
Haldane conjectured in 1983 that the half-odd integer and integer spins might behave
essentially differently \cite{hal}, which together with a field theoretic prediction of
Affleck \cite{aff} for a logarithmic correction to the power-law
behavior in the spin-spin correlation function
in the spin-$\frac{1}{2}$ case, triggered an intensive study of HA ranging from 
the exact diagonalization \cite{bot,par} and Monte Carlo \cite{nig,san} to 
DMRG (density matrix renormalization group) \cite{whi1,whi2,hall}.  These studies 
along with the Bethe ansatz solution for the spin-$\frac{1}{2}$ case \cite{dec} lead to
a confirmation of the both claims.  Concerning the elementary excitation for the entire Brillouin zone, however, 
Takahashi's two attempts following the Feynman variational method for $^4\mathrm{He}$ and 
a projector-Monte Carlo method were the only studies \cite{tak1,tak2}. And none of the previous studies gave a 
serious consideration to the expression (\ref{eq1}). 
   
In this Letter, we analyze (\ref{eq1}) exactly
for the HA (\ref{eq2}) with periodic boundary conditions 
by the recently developed entanglement perturbation theory (EPT).
EPT is a novel many-body method which takes into account correlations systematically. 
Its mathematical implementation is singular value decomposition (SVD),
 intuitively 
\textit{divide and conquer}. 
EPT has addressed so far classical statistical mechanics \cite{chu1}, 
1D quantum ground states \cite{chu2} and 2D quantum ground states \cite{chu3}.  
We here address 
 the elementary excitation in one dimension. 
By EPT, we are not only free from a negative sign problem which is inherent to
MC for quantum spins and fermions, but can also handle an order of magnitude larger systems than MC and DMRG.
The
key of the success lies in our ability of calculating the ground state $|g\rangle$ precisely and most
importantly in an \textit{un-renormalized} form.  
We examine the cluster excitation operator $O_l$ systematically.  
We have found that the size of the magnon is 4 lattices long at largest for both spin-$\frac{1}{2}$ and spin-1.

We solve the problem (\ref{eq1}) in two steps.  First, we find the ground state.  We can do this in two ways.
Either to consider the density matrix eigenvalue problem 
$e^{-\beta H}|g\rangle=e^{-\beta E_g}|g\rangle$ with $\beta \rightarrow 0$
as done in \cite{chu2} (EPT-g1). There, the density matrix is expressed in a matrix, tensor product
form, reducing the problem to that of the partition function calculation in the
2D, 3D Ising model \cite{chu1}. This method is particularly suited to study the
infinite system because then we only need to consider the largest eigenvalue.
 Or to directly minimize the energy $\langle g|H|g\rangle
/\langle g|g\rangle$ as formulated in 
 \cite{chu3} (EPT-g2).
The second method is free from the parameter $\beta$, but it needs to handle a number of 
eigenvalues no matter how large the system size.  
Either way, the starting wave function is that of SVD-ed one \cite{chu1,chu2,chu3},
\begin{equation}\label{eq3}
|g\rangle = \cdots \zeta^1_{\alpha \beta} (s_1)\zeta^2_{\beta \gamma} 
(s_2)\zeta^1_{\gamma \delta} (s_3)\cdots
\end{equation}
on the basis $\cdots |s_1 \rangle \otimes |s_2 \rangle \otimes |s_3 \rangle \otimes \cdots$
where $|s_i \rangle$ denotes spin states on the site $i$, and the local bi-partite wave functions
 $\zeta^{1,2}$ reflect the local antiferromagnetic interaction.  

\begin{figure}
\label{fig1}
\epsfig{file=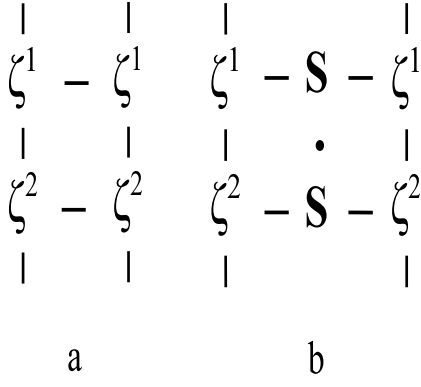,width=5.5cm,height=5.0cm}
\caption{
(a) Matrix $A$. The vertical lines denote entanglement while the 
horizontal ones the basis spin states. The upper 2 external verticle lines 
constitute the right index of $A$ while the lower 2 the
left index. The inner lines should be contracted. (b) Matrix $\hat{A}$
}
\end{figure}

Let us consider the variational problem, EPT-g2,
\begin{equation}\label{eq4}
\delta E_g=\delta \frac{\langle g| H | g \rangle}{\langle g | g \rangle}=0
\end{equation}
with respect to $\zeta^{1,2}$.
First consider $\delta \langle g | g \rangle$. 
For the periodic ring 
of $N$ bi-lattice units, we have 
\begin{equation} \label{eq5}
\langle g | g \rangle=T_r A^N
\end{equation}
with the matrix $A$ shown in Fig.1a. The variation with respect to $\zeta^{1,2}$ is, 
due to the translational symmetry, $N$ times the variation of $\zeta^{1,2}$ at
a particular bi-lattice unit.  The idea is to use an initial trial for $\zeta^{1,2}$
 and evaluate the trace over the $N-1$ bi-lattice units, which can be done as follows.
Writing the left and right eigenvector matrixes as $L$ and $R$ and the diagonal
eigenvalue matrix as $\sigma$, the matrix $A$ is written as 
\begin{equation} \label{eq6}
A=R\sigma \tilde{L}
\end{equation}
where $\tilde{L}$ is the transpose of $L$,
with the property $\tilde{L}\cdot R=1$. We can thus write
\begin{equation} \label{eq7}
\langle g | g \rangle = \sum_{i} \sigma_i^{N-1} \tilde{L_i} A R_i
\end{equation}
Now note that the last form (\ref{eq7}) contains $\zeta^{1,2}$ 
quadratically,
 thus written as either $\tilde{\zeta^1} M^1(\zeta) \zeta^1$ or 
$\tilde{\zeta^2} M^2(\zeta) \zeta^2$  
with
\begin{eqnarray}\label{eq8}
M^1(\zeta)_{l,r}=\delta_{ls,rs} \sigma_i^{N-1} L_i(lt,rt) R_i(lu,ru)
\nonumber\\
\zeta^2_{ld,lt}(ts)\zeta^2_{rd,rt}(ts)
\end{eqnarray}
where the indexes are put together as 
$l=(lu-1)\cdot p \cdot n_s+(ld-1)\cdot n_s+ls$ and 
$r=(ru-1)\cdot p \cdot n_s+(rd-1)\cdot n_s+rs$ 
where $p$ denotes the entanglement size 
and $n_s$ the local spin degrees of freedom, and summations are 
implied for the repeated indexes.  And likewise for $M^2(\zeta)$.
As for the numerator $\langle g |H| g \rangle$, 
because of the translational symmetry, the variation with respect to $\zeta^{1,2}$ is again $N$
times the variation of $\zeta^{1,2}$ at a particular bi-lattice unit. Since the Hamiltonian 
is a sum of local nearest neighbor interactions, a typical term in $\langle g |H| g \rangle $ is
$\tilde{L_j} A R_i \sigma_j^{l-2} \tilde{L_i} \hat{A} R_j \sigma_i^{N-l}$
where $\hat{A}$ is given by Fig.1b.  A summation is carried out over the terms containing
$\zeta^{1,2}$ except a quadratic term to be variated, and $\langle g |H| g \rangle$ can be 
written as either $\tilde{\zeta^1} N^1(\zeta) \zeta^1$ or $\tilde{\zeta^2} N^2(\zeta) \zeta^2$.
Thus the variational problem (\ref{eq4}) leads to nonlinear, generalized eigenvalue problems
\begin{table}
\begin{center}
\begin{tabular}{|c|c|c|c|}
  \hline
   size  & EPT & BA \\
  \hline
  16 & -0.4463935 & -0.4463935 \\
  \hline
  64 & -0.4433459 & -0.4433485 \\
  \hline
  256 & -0.4431555 & -0.4431597 \\
  \hline   
\end{tabular}
\caption{\label{table:ground}Comparison of the ground state energies between BA and EPT
with the entanglement $p=15$ for 16 spins, 22 for 64 and 36 for 256}
\end{center}
\end{table}
\begin{table}
\begin{center}
\begin{tabular}{|c|c|c|c|}
  \hline
  distance  & EPT & BA  \\
  \hline
  1 & -0.1477187 & -0.1477157 \\
  \hline
  2 & 0.0606790 & 0.0606798 \\
  \hline
  3 & -0.0502424 & -0.0502486 \\
  \hline
  4 & 0.0346217 & 0.0346528 \\
  \hline
  5 &  -0.0308335 & -0.0308904 \\
  \hline
  6 & 0.0243932 & 0.0244467 \\
  \hline
  7 & -0.0224726 & -0.0224982 \\
  \hline 
\end{tabular}
\caption{\label{table:corr}Comparison of $G^z$ 
over the first 7 sites between BA and EPT with $p=36$ for 256 spins}
\end{center}
\end{table}
\begin{equation}\label{eq9}
M^i(\zeta)\zeta^i=E_g N^i(\zeta)\zeta^i~~~;~~~i=1,2
\end{equation}
We have solved (\ref{eq9}) iteratively for both spin-$\frac{1}{2}$ and spin-1 for the system size
up to 1024 and the entanglement size up to 38.  A thorough discussion of the EPT result for the
ground state properties of the xxz model 
with the Ising anisotropy coupling $\lambda =0 \sim \infty$ 
will be given in a future publication \cite{lih}. We here concentrate
on the isotropic xxx case.
A convergence in the ground state energy per site occurs typically 
around p=20.  
\begin{figure}
\label{fig2}
\epsfig{file=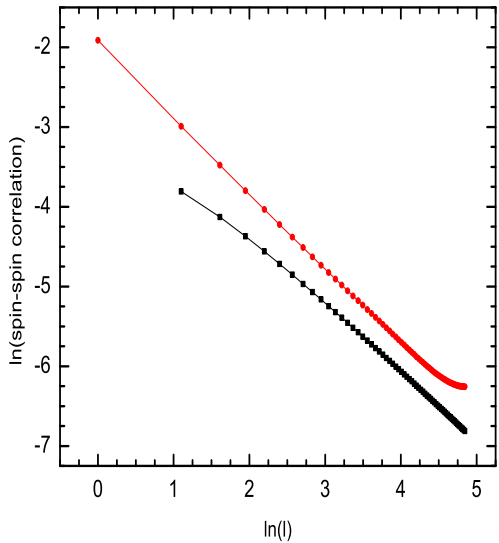,width=8.5cm,height=8.5cm}
\caption{
Log-log plot of the spin-spin correlation function
for 256 spins for spin-$\frac{1}{2}$. The upper line is 
the converged result for 
entanglement 32, while the lower line 
is the asymptotic formula from field theory.
}
\end{figure}

For spin-$\frac{1}{2}$, the exact result is given by Bethe ansatz (BA) \cite{kar}.  
The ground state energies per site for 16, 64 and 256 spins by EPT and BA
are compared in Table I. 
The spin-spin correlation
function $G^z(l)=\langle S_z(0)S_z(l) \rangle$ is also known by BA up to lattice separations 7.  
Comparison of BA and EPT for 256 spins is given in Table II.
As for a longer distance $l\gg 1$, an asymptotic formula is given by field theory \cite{aff}.
Extensive finite size analysis with exact diagonalization, Monte Carlo and DMRG \cite{bot,san,hall}
have been done concerning the field theoretic prediction of the logarithmic correction
of the form $\propto \sqrt{ln(l)}/l$ at large distance.  The lattice size
at which the asymptotic formula is realized  
is estimated to be several thousands.  Fig.2 shows the EPT result of the correlation function for 
the spin-$\frac{1}{2}$ with 256 spins which is a lot longer than previously studied.
Note that the converged and hence exact result at $p=32$ is well fitted by a straight line, 
except near the central part at $l=128$ reflecting the periodic boundary condition, 
\begin{equation}\label{eq10}
G^z(l) \varpropto l^{-0.92}
\end{equation}
While the greater than -1 exponent reflects the logarithmic correction, a large discrepancy between the 
EPT and the field theoretic asymptotic form, FT in Fig.2,
for a distance $l\sim100$ is consistent with the argument that the FT asymptotic form
 is correct for 
distances as large as several thousands.
It should be
noted that EPT can handle any system size, but for a larger system, 
a larger entanglement is necessary to get a good ground state.  An interesting question then is,
to calculate the correlation functions correctly for thousands-long distances, 
how large the entanglement $p$ should be?  In fact the issue is closely related to 
another field theoretic prediction \cite{aff} that the correlation functions 
at large distance
should have a singular dependence on the Ising anisotropy parameter $\lambda$,
namely $G^z=G^x$ at $\lambda=1$ but $G^z=4G^x$ when $\lambda$ approaches 1 from below. 
Following the successful EPT analysis of such symmetry breaking
in the 
2D, 3D Ising models \cite{chu1}, a calculation is currently underway and will be
reported elsewhere.

As for spin-1, EPT agrees with DMRG, e.g. the ground state energy per site for $N=48$ is
-1.401482 (EPT with $p=24$) vs -1.401484 (DMRG \cite{whi1}), and the spin-spin correlation function shows an
exponential decay as expected for a gapped system. 

An important note on the ground state algorithms EPT-g1,2 is that they 
can also calculate some excited states.  
For example, we have applied EPT-g2 to spin-1
to get the energy gap at $k=\pi$ correctly, 0.4124 (EPT with $p=20$) vs 0.4123 (DMRG \cite{whi1}) for 48 spins. 

We now come to the second step, the implementation of the variational program
for the elementary excitation (EPT-e),
\begin{equation}\label{eq12}
\delta E_k=\delta \frac{\langle \Psi_k|H|\Psi_k \rangle}{\langle \Psi_k|\Psi_k \rangle}=0
\end{equation}
Using (\ref{eq1}), we first rewrite $E_k$ as 
\begin{equation}\label{eq13}
\varepsilon(k)=E_k-E_g=\sum_{l=1}^{2N}e^{-ikl} \frac{\langle g | O_l [H,O_0^{\dagger}] |g \rangle}
{\langle \Psi_k | \Psi_k \rangle}
\end{equation}
where $[\cdots]$ means a commutator and we have used the fact that $H|g \rangle = E_g |g \rangle$,
and therefore the accurate ground state wave function is a crucial ingredient in this method.
Note that $N$ is the number of bi-lattice units and $2N$ is the total number of spins.
More importantly, we need the \textit{un-renormalized} ground state
 which may not be easy to
obtain by DMRG without an efficient restoration of the un-renormalized ground
state at the end of the calculation, particularly for large systems such as 512 and 1024 spins
as studied by EPT.

To carry out the variation $\delta \varepsilon (k)=0$ with respect to the cluster operator $O_l$,
the simplest case is the one which acts only on one site and, say,
for the spin-triplet excitation, then it is uniquely $S^+=S_x+i S_y$ for spin-$\frac{1}{2}$.  
In the spin-1 case,
there are only two such operators, $\mu_1=S^+$ and $\mu_2=S^+S_z$.
The cluster operator can then be written as a 
linear combination $O_l=c_1 \mu_1 +c_2\mu_2$ and the variation is
with respect to the vector $\tilde{x}=(c_1,c_2)$, leading to
a generalized eigenvalue problem
\begin{equation}\label{eq14}
Tx=\varepsilon(k)Ux
\end{equation}
where T and U are $2 \times 2$ matrixes. 
The n-cluster operator $O_l$ is generally written as a linear combination of operator
products of n $n_s \times n_s$ local operators, where $n_s=2$ for spin-$\frac{1}{2}$ and 3 for
spin-1.  
With the increase of the cluster size of
$O_l$, the matrix size of T and U increases like 1,4,15 and 56 (spin-$\frac{1}{2}$)
and 2,16,126 and 1016 (spin-1) for the spin-triplet excitation.
For example, for $n=2$ and spin-$\frac{1}{2}$, 
$O_l$ is a linear combination of the 4 local excitation operators, 
$S^+ \otimes 1$, $S^+ \otimes S_z$,
$1 \otimes S^+$ and $S_z \otimes S^+$.
The calculation of the matrixes T and U are essentially the same as in the ground state,
although due to the \textit{cluster} nature of the operator $O_l$ and the presence of the
commutator $[H,O_0^{\dagger}]$, we have a little lengthy algebraic procedure,
 details to be presented elsewhere.  

Fig.3 shows the spin-triplet excitation spectrum for 
spin-$\frac{1}{2}$ with the chain size 512 and the cluster size $n$ up to 4.  
The EPT calculation almost converged at $n=4$,
and gives an agreement of $1\%$ precision with BA \cite{dec}.  Fig.4 shows the same for spin-1
up to the cluster size $n=3$ where the calculation almost converged.  The 
 Haldane gap at $k=\pi$ is found by EPT to be 0.414 agreeing with the previous results \cite{nig,tak2,whi2}.  
As for the region $0\leqslant k/\pi \leqslant 1/2$, the spin-triplet spectrum is believed to be embedded
in a continuum spectrum of a pair of magnons with the total spin-z component to be 0 \cite{ma}.  The exact diagonalization
for $N=14$ indeed tells us that the lowest excitation at $k=0$ is spin singlet, presumably a $(-\pi,\pi)$ pair of 
spin-triplet magnons from $k=\pm \pi$ \cite{par,tak2}. We believe that the spin-triplet magnon spectrum for 
spin-1 for the entire Brillouin zone has been determined for the first time by EPT.  
Moreover EPT can handle not only the lowest but the entire spin-triplet sectors, and can be repeated for other
excitations such as spin-singlet. It simply amounts to calculating not just the minimum but all the eigenvalues of 
(\ref{eq14}).  Such entire excitation spectra have been known only for the isotropic spin-$\frac{1}{2}$
case by Bethe ansatz, see Fig.4, 5 in \cite{kar}.
\begin{figure}
\label{fig3}
\epsfig{file=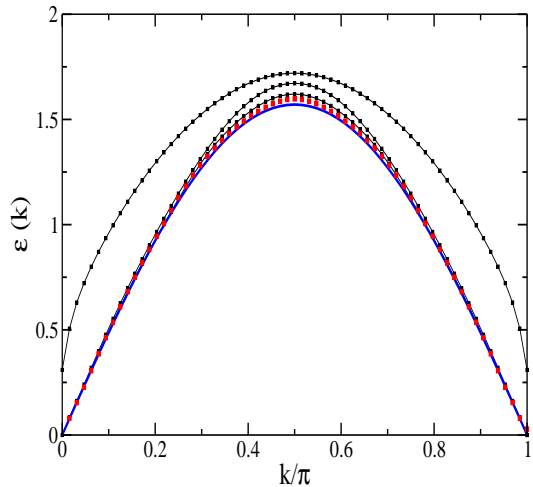,width=8.0cm,height=8.0cm,angle=-90}
\caption{
Spin triplet excitation spectrum for spin-$\frac{1}{2}$ for 512 spins. The ground state
used is from p=20.  From top to bottom, the cluster size is 
1 to 4 (red online). The thick line (blue online) is the Bethe ansatz result \cite{dec}.
}
\end{figure}

In conclusion, we have developed EPT for the elementary excitation (\ref{eq1}),
 where
the \textit{un-renormalized} ground state $|g \rangle$ plays a central role.
A challenge to EPT is to calculate the correlation functions for a distance
long enough to compare with field theory.
While the applications of EPT-e to 1D fermions and bosons are straightforward,  
we need yet to see how it works in two dimensions. 
Finally, the successful calculation
of the excitation spectrum indicates that EPT can handle various
nano-structures embedded
in \textit{correlated}
 host materials, opening a possible new look at the Kondo effect \cite{ott}. 

\begin{acknowledgments}
This work was partially supported by the NSF under grant No.PHY060012N
and utilized the TeraGrid Cobalt at NCSA
at UIUC. This work also partially utilized 
the College of Sciences and Humanities Cluster at the Ball State University. 
\end{acknowledgments}
\begin{figure}
\label{fig4}
\epsfig{file=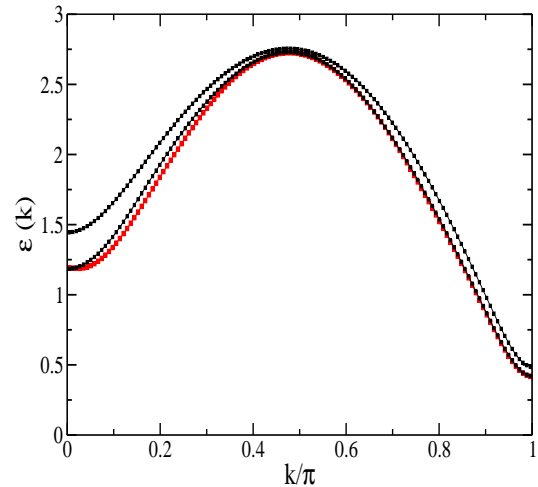,width=8.0cm,height=8.0cm,angle=-90}
\caption{
Spin triplet excitation spectrum for spin-1 for 512 spins. The ground state
used is from p=24.  From top to bottom, the cluster size is 
1 to 3 (red online). 
}
\end{figure}
\bibliography{excitation}

\end{document}